\documentclass[runningheads]{casc}

\font\dunh=cmdunh10 scaled 1000
\font\smallrm=cmr7

\def\EinS{{\dunh EinS}}
\def\eins{{\rm E{\smallrm in}\kern-.07em\rm S}}

\def\Mathematica{{\it Mathematica}}

\begin{document}

\pagestyle{headings}

\mainmatter

\title{
Independent Components
of an Indexed Object
\\with Linear Symmetries}

\titlerunning{Independent Components
of an Indexed Object}

\author{Sergei A. Klioner}

\authorrunning{Sergei A. Klioner}

\institute{
Lohrmann Observatory, Dresden Technical University,\\
01062 Dresden, Germany\\
\email{klioner@rcs.urz.tu-dresden.de}
}

\maketitle

\begin{abstract}
The problem of finding independent components of an indexed object
(e.g., a tensor) with arbitrary number of indices and arbitrary linear
symmetries is discussed. It is proved that the number of independent
components $f(k)$ is a polynomial of
degree not greater than the number of indices $n$,
$k$ being the dimension of the space. Several algorithms to
compute $f(k)$ for arbitrary $k$ are described and discussed. It is
shown that in the worst case finding $f(k)$ for arbitrary $k$ requires
solving at most $P(n)$ systems of linear equations with at most
$(n!)^2$ equations for at most of $n!$ unknowns, $P(n)$ being the
number of partitions of $n$. As a by-product, an efficient algorithm to
parametrize all components of the object through its independent
components is found and implemented in \Mathematica.
\end{abstract}

\section{Motivation}

In this paper we attempt to find an approach to the problem formulated,
e.g., in \cite{Hartley:1996}. The problem consists in calculating the
number of independent components of an indexed object $A^{ijk\dots}$
with symmetries (for example, $A^{ijk\dots}=A^{kij\dots}$). The
importance of this problem for computer algebra systems for indicial
tensor computations can be clarified by the following example given in
\cite{Hartley:1996}. In many packages for indicial tensorial
calculations one can define an object $A^{ijk}$ with 3 indices and the
following two symmetry properties:
\begin{equation}
A^{ijk}=A^{jik}
\nonumber
\end{equation}

\noindent
and
\begin{equation}
A^{ijk}=-A^{ikj}.
\nonumber
\end{equation}

\noindent
One can easily check that these two symmetry properties imply that all
components of $A^{ijk}$ is zero. However, none of the packages
available to the author can automatically recognize this fact. In many
cases, however, if the software could recognize such and similar
situations, it could simplify many kinds of calculations. On the other
hand, it is often quite important to find not only the number of
independent components, but also the independent components themselves (for
example, to store the components of an object in the best possible
way). In this paper we will give efficient algorithms for both
finding and counting the independent components of an object with
symmetries.

\section{The Problem and Some Trivial Properties}

Here the following problem with linear symmetries is considered. Let
$A^{i_1i_2\dots i_n}$ be an object with $n$ indices. Each index $i_j$
can take values from $i_j\in\{1,2,\dots,k\}$, where $k\ge1$ is the
dimensionality of the space where the object is defined. The object has
$s$ symmetries defined by equations of the form
\begin{equation}
\label{symmetries}
\sum_{j=1}^{p_l} a_{jl}\ A^{\pi_{jl}}=b_l, \qquad l=1, 2, \dots, s
\end{equation}

\noindent
where $a_{jl}$ and $b_l$ are numbers characterizing the $l$-th
symmetry, and $\pi_{jl}$ is a permutation of the $n$ indices $i_1i_2\dots
i_n$, $p_l$ is the number of terms in the $l$-th symmetry. Clearly, the
maximal number of the terms in one symmetry and the maximal number of
linearly independent symmetries are constrained as (otherwise the
individual terms and the whole symmetries linearly depend on the
other terms and symmetries, respectively)
\begin{eqnarray}
p_l\le n!\ ,
\\
s\le n!\ .
\end{eqnarray}

\noindent
In principle, the number of symmetries $s$ can be larger than $n!$, but
in this case either the additional symmetries are linearly dependent on
the first $n!$ symmetries or at least one of the additional symmetries
are incompatible with the first $n!$ ones and such an object does not
exist. A component of $A^{i_1i_2\dots i_n}$ is called independent if it
is not constrained by symmetries (\ref{symmetries}) to be a number and
if it cannot be calculated as a function of other independent
components. The set of independent components is obviously not unique
(if a symmetry requires $A^{21}=A^{12}$ any of these two components can
be considered as independent). The size of this set is, however,
independent of this non-uniqueness, and it is sufficient to find any
version of it. The problem consists in computing of the number $f$ of
independent components of $A^{i_1i_2\dots i_n}$ as a function of $k$:
$f(k)$. Since the total number of components of $A^{i_1i_2\dots i_n}$
for a fixed $k$ is $k^n$, one concludes
\begin{equation}
\label{f(k)-upper}
0\le f(k)\le k^n.
\end{equation}

\section{Objects with One and Two Indices}

In order to understand the problem more clearly, let us consider first
all possible objects with one and two indices. The case of $n=1$ is quite trivial.
An object $A^i$ has at most $k$ components. If the object has no symmetries
of the form (\ref{symmetries}), the number of independent components $f(k)=k$.
The only possible symmetry reads
\begin{equation}
\label{symmetry-1}
a\, A^i = b,
\end{equation}

\noindent
where $a$ and $b$ cannot vanish simultaneously (otherwise the object
has again no symmetry). Clearly, if $a=0$ and $b\neq0$ the symmetry
cannot be satisfied. Such symmetries will be called incompatible. In
case of incompatible symmetries we will symbolically write
$f(k)=\emptyset$. If $a\neq0$ then $A^i=b/a$, i.e. all components for
any $k$ are constrained to be number $b/a$. This can be summarized as
follows
\begin{eqnarray}
\label{f(k)-1}
f(k) = \left[\ \ \begin{array}{ll}
               k,\ &\ {\rm no\ symmetries}\ (a=b=0),\\
               0,\ &\ a\neq0,\\
               \emptyset,\ &\ a=0\ \&\ b\neq 0.
              \end{array}
       \right.
\end{eqnarray}

The case of $n=2$ is a bit less trivial. For an object $A^{ij}$ with no
symmetries $f(k)=k^2$. Each of the two possible symmetries is of the form
\begin{equation}
\label{symmetry-2}
a_1 A^{ij}+a_2 A^{ji} = b.
\end{equation}

\noindent
Let us consider first only one symmetry (\ref{symmetry-2}). For $C_k^1=k$
``diagonal'' components with $j=i$ one can write this symmetry as (no
implicit summation is assumed here)
\begin{equation}
\label{symmetry-2-diagonal}
(a_1+a_2) A^{ii} = b, \quad i=1,2,\dots,k\,.
\end{equation}

\noindent
This symmetry can be analyzed in the same way as we did for
(\ref{symmetry-1}) above. The $k$ ``diagonal'' components are
constrained independently of each other (but clearly in the same way).
Therefore, it is sufficient to
analyze (\ref{symmetry-2-diagonal}) for one fixed $i$ and then multiply
the result (either 0 or 1 independent components) by $k$. Among the
other $k(k-1)$ ``off-diagonal'' components with $j\neq i$ there exist
$C_k^2$ pairs of components, $C_a^b=\displaystyle{a!\over b!(a-b)!}$
being the binomial coefficient. It is only the components within each
of these pairs which could be potentially constrained by
(\ref{symmetry-2}). Indeed, for any fixed $j$ and $i$ ($j\neq i$) only
two combinations of indices, $ij$ and $ji$, are connected to each other
by a permutation and appear in (\ref{symmetry-2}). Therefore, it is
sufficient to consider any of these pairs and multiply the results (one
can have 0, 1 or 2 independent components within each pair) by
$C_k^2=k(k-1)/2$. Therefore, the number of independent components of
$A^{ij}$ for an arbitrary $k$ can always be calculated as
$f(k)=f_2\,k\,(k-1)/2 +f_1\,k$ with two integers $0\le f_2\le2$ and
$0\le f_1\le1$ depending on the symmetries of the object. Let $i=1$ and
$j=2$. Then (\ref{symmetry-2}) give two linear equations constraining
$A^{12}$ and $A^{21}$:
\begin{eqnarray}
\label{symmetry-2-exp}
a_1 A^{12}+a_2 A^{21} &=& b,
\nonumber\\
a_1 A^{21}+a_2 A^{12} &=& b.
\end{eqnarray}

\noindent
Analyzing (\ref{symmetry-2-diagonal}) and (\ref{symmetry-2-exp}) one gets
the number of independent components of an object with $n=2$ and
one symmetry (\ref{symmetry-2}):
\begin{equation}
\label{f(k)-2}
f(k) = \left[\ \ \begin{array}{ll}
               k^2,\ &\ {\rm no\ symmetries}\ (a_1=a_2=b=0),\\[3pt]
               {1\over 2}\,k\,(k+1),\ &\ a_1=-a_2\neq0\ \&\ b=0,\\[3pt]
               {1\over 2}\,k\,(k-1),\ &\ a_1=a_2\neq0,\\[3pt]
               0,\ &\ |a_1|\neq|a_2|,\\[3pt]
               \emptyset,\ &\ a_1+a_2=0\ \&\ b\neq 0.
                 \end{array}
       \right.
\end{equation}

\noindent
Clearly, for $a_1=-a_2\neq0\ \&\ b=0$ the symmetry (\ref{symmetry-2})
can be written as $A^{ij}-A^{ji}=0$. This means that the ``diagonal''
components are not constrained at all, and the ``off-diagonal''
components are pairwise equal. We have thus a symmetric ``matrix''. For
$a_1=a_2\neq0$ one has $A^{ij}+A^{ji}=\overline{b}=b/a_1$. This means
that the ``diagonal'' components are all equal to $\overline{b}/2$ and the
``above-diagonal'' components can be computed from the
``below-diagonal'' ones as $A^{ij}=-A^{ji}+\overline{b}$. For
$\overline{b}=0$ we have a skew-symmetric ``matrix'' here. In case
$|a_1|\neq|a_2|$ all components are equal to the same number:
$A^{ij}=b/(a_1+a_2)$.

It is easy to see that for objects $A^{ij}$ with two symmetries
\begin{eqnarray}
\label{symmetry-3}
a_{11} A^{ij}+a_{21} A^{ji} &=& b_1,
\nonumber\\
a_{12} A^{ij}+a_{22} A^{ji} &=& b_2
\end{eqnarray}

\noindent
$f(k)$ also takes the same 5 possible values as in (\ref{f(k)-2}).
%
%
The ideas found in these examples will allows us to formulate several
important theorems and algorithms below.

\section{The Number of the Independent Components}

Let us first leave aside the question whether the symmetries of the
object are compatible and consider that $f(k)\neq\emptyset$ for any
$k$. This will be further discussed in Section
\ref{Section-compatibility}. The results of the previous Section
suggest to formulate the following

\begin{theorem}
\label{Theorem-1}
For an object $A^{i_1i_2\dots i_n}$ with arbitrary $n$ and arbitrary
compatible linear symmetries the number of independent components $f(k)$ is a
polynomial of the dimensionality $k$. The degree of this polynomial does not
exceed $n$.
\end{theorem}

\begin{proof}
From (\ref{f(k)-upper}) it is obvious that if $f(k)$ is a polynomial
its degree cannot exceed $n$.

Let us denote the set of all possible values of each of the indices
$i_a$ as $N_k=\{1,2,\dots,k\}$. Its cardinality is $|N_k|=k$. Let us
also denote the set of all independent components of an object
$A^{i_1i_2\dots i_n}$ with all $i_a\in N_k$ as $E(N_k)$. The cardinality of
$E(N_k)$ is clearly $f(k)$. It is clear that we can change $N_k$ to any
set $S$ with the same cardinality $k$ and number of independent components
of $A^{i_1i_2\dots i_n}$ with $i_a\in S$ will be again $f(k)$. That is,
for any $S$ one has $|E(S)|=f(|S|)$.

Let $S^a$, $1\le a\le k$ are the $C_k^{k-1}=k$ subsets of $N_k$ such
that $|S^a|=k-1$. Clearly, $S^1\cup S^2\cup\dots\cup S^k=N_k$. Any two
components can be potentially related to each other according to
(\ref{symmetries}) only if their sets of $n$ indices are related by a
permutation. Thus, if $k>n$ the set $E(N_k)$ of independent components
of $A^{i_1i_2\dots i_n}$ with $i_a\in N_k$ is the union of the $k$ sets
$E(S^a)$ of independent components with $i_a\in S^a$:
$E(N_k)=E(S^1)\,\cup\,E(S^2)\,\cup\,\dots\,\cup\,E(S^k)$. For the same
reason for any two sets $S_1$ and $S_2$ one has $E(S_1)\cap
E(S_2)=E(S_1\cap S_2)$. Applying the Inclusion-Exclusion Principle
\cite{Comtet:1974,Jukna:2001} to the $k$ sets $E(S^a)$ one gets
\begin{eqnarray}
\label{inc-exc-principle}
|E(N_k)|=|\cup_{a=1}^k E(S^a)|&=&
\sum_{1\le a\le k} |E(S^a)|
\nonumber\\
&&
-\sum_{1\le a<b\le k} |E(S^a)\cap E(S^b)|
\nonumber\\
&&
+\sum_{1\le a<b<c\le k} |E(S^a)\cap E(S^b)\cap E(S^c)|
-\dots
\nonumber\\
&&
+(-1)^{k-1}\,|E(S^1)\cap \dots \cap E(S^k)|, \quad k>n\,.
\end{eqnarray}

\noindent
This formula immediately implies
\begin{equation}
\label{f(k)-via-f(k-1)}
f(k)=\sum_{j=1}^{k-1} (-1)^{j-1} f(k-j) C_k^j, \quad k>n\,.
\end{equation}

\noindent
Since this last formula is valid for any $k>n$ one can compute $f(k)$ for
$k>n$ from the $n$ values $f(i)$ for $i=1,\dots, n$. This means
that the $f(k)$ is a polynomial of $k$ of degree $n$ or less. \qed
\end{proof}

From the computational point of view it is useful to explicitly compute the
above-mentioned representation of $f(k)$ through the $n$ values $f(i)$
for $i=1,\dots, n$.

\begin{theorem}
\label{Theorem-poly-formula-n}
For an object $A^{i_1i_2\dots i_n}$ with arbitrary $n$ and arbitrary
compatible linear symmetries the number of independent components $f(k)$
for any $k>n$ can be computed as
\begin{equation}
\label{f(k)-f(1)-...-f(n)}
f(k)=\sum_{i=1}^n (-1)^{n-i}\,f(i)\,C_k^i\,C_{k-1-i}^{n-i}.
\end{equation}

\end{theorem}

\begin{proof}[1]
A straightforward way to prove (\ref{f(k)-f(1)-...-f(n)}) is by
induction. For $k=n+1$ Eq. (\ref{f(k)-f(1)-...-f(n)}) coincides with
(\ref{f(k)-via-f(k-1)}) and is thus correct. Suppose that Eq.
(\ref{f(k)-f(1)-...-f(n)}) is valid for some $k$. Let us prove that it is
also valid for $k+1$. Applying Eq. (\ref{f(k)-via-f(k-1)}) one gets
\begin{equation}
\label{f(k+1)-via-f(k)}
f(k+1)=\sum_{i=1}^{k} (-1)^{k-i}\, f(i)\, C_{k+1}^i.
\end{equation}

\noindent
Since for all $f(i)$ with $n+1\le i\le k$ Eq. (\ref{f(k)-f(1)-...-f(n)})
is supposed to be correct, one has
\begin{eqnarray}
\label{e1}
f(k+1)&=&\sum_{i=1}^{n} (-1)^{k-i}\,f(i)\,C_{k+1}^i
+\sum_{i=n+1}^{k} (-1)^{k-i}\,f(i)\,C_{k+1}^i
\nonumber\\
&=&
\sum_{i=1}^{n} (-1)^{k-i}\,f(i)\,C_{k+1}^i
+
\sum_{j=n+1}^k (-1)^{k-j}\,C_{k+1}^j
\sum_{i=1}^n (-1)^{n-i}\,f(i)\,C_j^i\,C_{j-1-i}^{n-i}
\nonumber\\
&=&
\sum_{i=1}^{n}(-1)^{n-i}\,f(i)\,
\left[
(-1)^{k-n}\,C_{k+1}^i
+\sum_{j=n+1}^k (-1)^{k-j}C_{k+1}^j\,C_j^i\,C_{j-1-i}^{n-i}
\right]
\nonumber\\
&=&
\sum_{i=1}^{n}(-1)^{n-i}\,f(i)\,C_{k+1}^i\,C_{k-i}^{n-i}.
\end{eqnarray}

\noindent
Here we used the well-known properties of the binomial coefficients
\cite{Comtet:1974,Jukna:2001}. In particular, we used
\begin{equation}
\label{binomial-identity}
\sum_{s=0}^{k-n} (-1)^s\,C_{k-i+1}^{n-i+1+s}\,C_{n-i+s}^s=1, \quad 0\le i\le n\le k.
\end{equation}

\noindent
Therefore, Eq. (\ref{f(k)-f(1)-...-f(n)})
is correct for any $k>n$. \qed
\end{proof}

\begin{proof}[2]
A more elegant proof of Eq. (\ref{f(k)-f(1)-...-f(n)}) directly
follows from Theorem~\ref{Theorem-1}: if a function is a polynomial
of degree $n$ or less, one can take its first $n$ values $f(i)$ for $i=1,2,\dots,n$
and construct a polynomial of degree $n$ having these values for $i=1,2,\dots,n$
(and $f(0)=0$) using the Lagrange interpolation formula. Indeed,
\begin{eqnarray}
\label{binomial-lagrange}
&&(-1)^{n-i}\,C_k^i\,C_{k-1-i}^{n-i}=
{
(k-0)(k-1)\dots(k-(i-1))(k-(i+1))\dots(k-n)
\over
(i-0)(i-1)\dots(i-(i-1))(i-(i+1))\dots(i-n)
}\ ,
\end{eqnarray}

\noindent
which is exactly the Lagrange form of the coefficients of the
interpolation polynomial.\qed
\end{proof}

\section{On the Compatibility of the Symmetry Properties}
\label{Section-compatibility}

Up to now we have ignored the question of compatibility of the
symmetry properties (\ref{symmetries}). One can easily check if the symmetries
are compatible for $k=1$.

\begin{theorem}
For an object $A^{i_1i_2\dots i_n}$ with arbitrary $n$
its symmetry properties (\ref{symmetries}) are incompatible for
$k=1$ if and only if at least one of the two conditions are met:

\begin{itemize}
\item[(1)] for at least one symmetry $b_l\neq 0$ and $c_l=0$ with
\begin{equation}
\label{c_l}
c_l=\sum_{j=1}^{p_l} a_{jl},
\end{equation}

\item[(2)] there exist at least two symmetries for which
$c_l\neq0$ and $c_{l^\prime}\neq0$,
and $b_l/c_l\neq b_{l^\prime}/c_{l^\prime}$.
\end{itemize}

\end{theorem}

\begin{proof}
For $k=1$ any of the symmetries can be written as
\begin{equation}
\label{symmetry-n-1}
c_l\,A^{11\dots 1}=b_l,
\end{equation}

\noindent
$c_l$ being defined by (\ref{c_l}).
These properties can be analyzed in the same way as we did for (\ref{symmetry-1}).
The theorem immediately follows from this analysis.\qed
\end{proof}

Another two important results are:

\begin{theorem}
\label{Theorem-comp-if-K->K-1}
If symmetries of an object $A^{i_1i_2\dots i_n}$ with arbitrary $n$ are
compatible for dimension $k=K$, they are also compatible for any dimension
$k<K$.
\end{theorem}

\begin{proof}
This is obvious since the whole system of symmetry-induced equations
for any $k<K$ is a part of the corresponding system for $k=K$.\qed
\end{proof}

\begin{theorem}
\label{Theorem-comp-if-K=N->K}
If symmetries of an object $A^{i_1i_2\dots i_n}$ with arbitrary $n$ are
compatible for dimension $k=n$, they are also compatible for
any dimension $k$.
\end{theorem}

\begin{proof}
For $k<n$ this follows from Theorem \ref{Theorem-comp-if-K->K-1}
while for $k>n$ from Theorem \ref{Theorem-poly-formula-n}.\qed
\end{proof}

\noindent
The algorithmic usage of Theorem \ref{Theorem-comp-if-K=N->K}
is clear: one has to check if the symmetries are compatible
for $k=n$.

\section{Algorithms to Compute the Independent Components}
\label{Section:algorithms}

Below three different algorithms to find the independent
components of an indexed object are discussed.

\subsection{Algorithm A}

The first algorithm is a straightforward one. For a fixed $k$ all
possible combinations of the numerical values of all $n$ indices are
substituted into each of the symmetries (\ref{symmetries}). Thus, one
obtains a system of linear equations for all $k^n$ components of the
object. The total number of equations is $s\,k^n$, where $s$ is the
number of symmetries ($s\le n!$). Each equation involves at most $n!$
components. Then, the system is solved and the independent components
are found explicitly. Counting them allows one to get $f(k)$ for that
fixed $k$ for which the system was generated. It is clear that if $n$, $k$
or $s$ are large the calculations can be very time-consuming. The only
reason to implement such an algorithm is the possibility to check
better algorithms described below.

\subsection{Algorithm B}

From the algorithmic point of view, Eq. (\ref{f(k)-f(1)-...-f(n)})
allows one to calculate $f(k)$ for any $k>n$ as soon as one has
calculated $f(k)$ for all $1\le k\le n$. Moreover, having the set of the
independent components for $k=n$ one can count the number of components
among them with indices $i_a\in N_p$, $N_p=\{1,\dots,p\}$ for
$p=1,...,n-1$. The number of such components is exactly $f(p)$, and,
therefore, it is sufficient to have the set of independent components
for $k=n$ to compute $f(k)$ for any $k$. Such an algorithm requires
solving at most $s\,n^n$ linear equations with at most $n!$ unknowns in
each equation and at most $n^n$ unknowns in the whole system (note that
each of the $s\le n!$ symmetries (\ref{symmetries}) generates at most
$n^n$ linear equations for at most $n!$ unknowns). Although, this
algorithm is better than straightforward calculation of the independent components
for some large $k>n$, it is still quite time-consuming for
larger $n$.

Another point is that this algorithm does not allow to list the
independent components and to represent the other components as
functions of the independent ones for $k>n$. It is certainly possible
to augment the algorithm in this direction. However, attempts to do so
allowed the author to formulate much more efficient algorithm for both
computing $f(k)$ and finding the dependencies in explicit form. This
algorithm described in the next Section.

\subsection{Algorithm C}

Here we suggest a much faster algorithm based on the fact that the
sequences of indices of the components which could be potentially
constrained by a symmetry of the form (\ref{symmetries}) are related
with each other by a permutation. This obvious fact has been already
mentioned and used above. Therefore, we can split the whole set of
$k^n$ components into such subsets within which the sequences of
indices are related by a permutation and then generate and solve the
corresponding linear equations only for the components within each of
these subsets. Moreover, one can drastically reduce the number of the
subsets to be considered since many of them are similar to each other
(e.g., they can be obtained from each other by changing, say, value 4
for all indices into value 5).

Let us consider a component $A^{i_1i_2\dots i_n}$ with some fixed indices
$1\le i_a\le k$. Any sequence of indices $i_1i_2\dots i_n$ can be
characterized by sequence $X=(x_1,\dots,x_k)$, where each $x_b$ is the
number of such $i_a$ in $i_1i_2\dots i_n$ that $i_a=b$. Clearly, one
has $x_1+x_2+\dots+x_k=n$ with constrains $0\le x_b\le n$. For fixed
$n$ and $k$ there are $C_{k+n-1}^n$ different solutions of this equation
with these constrains, and, therefore, $C_{k+n-1}^n$ different
sequences $X$.

It is clear that two components $A^{i_1i_2\dots
i_n}$ and $A^{j_1j_2\dots j_n}$  can be related to each other by a
symmetry of the form (\ref{symmetries}), only if both sequences of indices
$i_1i_2\dots i_n$ and $j_1j_2\dots j_n$ correspond to the one and same
sequence $X$. The number of components $A^{i_1i_2\dots i_n}$
corresponding to the same $X$ is the multinomial coefficient
$(n;x_1,x_2,\dots,x_k)=n!/(x_1!x_2!\dots x_k!)$ with
$x_1+x_2+\dots+x_k=n$.

The subsets of the components corresponding to two different $X_1$ and
$X_2$ can be treated in the same way if $X_1$ and $X_2$ are related to
each other by a permutation. Such a permutation corresponds just to
renaming all indices having, say, value 1 to, say, 5, and so on. The
two subsets corresponding to two such $X_1$ and $X_2$ have the same
number of independent components and the same dependence of the
other components on the independent ones. Therefore, it is
sufficient to calculate the dependence of the components only for one
$X$ among all of them related to each other by a permutation. We will
consider only the sorted version $Y$ of $X$: $Y=(y_1,\dots,y_k)$ with
$y_1\ge y_2\ge \dots\ge y_k$.

Let $Y^l$ be the sequence $Y$ with exactly $l$ nonzero elements:
$Y^l=(y_1,\dots,y_l,0,\dots,0)$, $|Y^l|=k$, $y_1\ge y_2\ge\dots\ge
y_l>0$. It is obvious that $1\le l\le \min{(n,k)}$. Each $Y^l$
corresponds to a partition of $n$ into $l$ parts. There exists $P(n,l)$
different partitions of this kind.

Now, let $p\le l$ denote the number of distinct values among $y_1,y_2,\dots,
y_l$, and $1\le s_m\le n$, $1\le m\le p$ is how many times the value number
$m$ (which is one of the $p$ distinct values in $Y^l$) appears among
$y_1,y_2,\dots, y_l$. Then the number of different $X$ which can be obtained
by permutations from $Y^l$ is $C_k^l\,l!/(s_1!s_2!\dots s_p!)$.

Combining all the results discussed above one gets the following formula
for the number of independent components
\begin{equation}
\label{algorithm-2}
f(k)=\sum_{l=1}^{\min{(n,k)}}\left[\sum_{j=1}^{P(n,l)} g(Y^l_j)\,{l!\over s_1^j!s_2^j!\dots s_p^j!}\,\right]\,C_k^l,
\end{equation}

\noindent
where the inner sum goes over all the partitions of $n$ into $l$ parts,
each such partition corresponds to $Y^l_j$, the numbers $s_1^j$, $s_2^j$,\dots
$s_p^j$ is calculated for $Y^l_j$, and $g(Y^l_j)$ is the number of
independent components among $n!/(y_1!\dots y_l!)$ components
$A^{i_1i_2\dots i_n}$ corresponding to $Y^l_j$:
\begin{equation}
\label{algorithm-2-g}
0\le g(Y^l_j)\le {n!\over y_1!\dots y_l!}.
\end{equation}

\noindent
Two identities can be used to check the internal consistency of
(\ref{algorithm-2}) and (\ref{algorithm-2-g}). First, the total number
of sequences $X$ for fixed $n$ and $k$
can be calculated from (\ref{algorithm-2}) with $g(Y^l_j)=1$ and
should be $C_{k+n-1}^n$ as discussed above. One can see that this is true:
\begin{equation}
\label{algorithm-2-check-1}
\sum_{l=1}^{\min{(n,k)}}\left[\sum_{j=1}^{P(n,l)} {l!\over s_1^j!\dots s_p^j!}\,\right]\,C_k^l=C_{k+n-1}^n.
\end{equation}

\noindent
Second, for an object without symmetries $g(Y^l_j)=n!/(y_1!\dots
y_l!)$, and the total number of components
 for fixed $k$ and $n$ calculated
according to (\ref{algorithm-2}) should be $k^n$. Indeed, it is also
true:
\begin{equation}
\label{algorithm-2-check-2}
\sum_{l=1}^{\min{(n,k)}}\left[\sum_{j=1}^{P(n,l)} {n!\over y_1!\dots y_l!}\,{l!\over s_1^j!\dots s_p^j!}\,\right]\,C_k^l=k^n.
\end{equation}

Combining (\ref{algorithm-2}) and (\ref{f(k)-f(1)-...-f(n)}) it is clear that
in order to calculate $f(k)$ for arbitrary $k$ it is sufficient to calculate
the number of independent components within the subsets of $A^{i_1i_2\dots
i_n}$ corresponding to $\sum_{l=1}^n P(n,l)= P(n)$ different sequences
$Y^l_j$, $P(n)$ being the total number of partitions of $n$. The size of the
subsets of $A^{i_1i_2\dots i_n}$ does not exceed $n!$. Therefore, in the
worst case one should solve $(n!)^2$ linear equations (each symmetry
(\ref{symmetries}) generates at most $n!$ distinct equations, and there are
at most n! symmetries) with $n!$ unknowns. This is much better that for the
algorithms A and B.

\subsection{Reduction of the Number of Linear Equations}
\label{Section:reduction}

A simple idea allows one to reduce further the number of
equations in the system generated by the symmetries before solving that system
of these equation. One can
put the equations into a canonical form in which it is trivial to check
if any two equations are equivalent (e.g., the numerical coefficient at
the lexicographically first component should be equal to unity) and
retain only one among the equivalent equations
possibly appearing in the set of generated equations. For example, the
symmetry $T^{ij}-T^{ji}=0$ for $i\in\{1,2\}$ produces two equations
$T^{12}-T^{21}=0$ and $T^{21}-T^{12}=0$ which are equivalent and can be
considered as one equation. On the other hand, the symmetry $3\, T^{ij}+4\,
T^{ji}=7$ for $i\in\{1,2\}$ gives two linearly independent equations $3\,
T^{12}+4\, T^{21}=7$ and $3\, T^{21}+4\, T^{12}=7$ whose canonical forms are
different: $T^{12}+{4\over3}\, T^{21}={7\over3}$ and $T^{12}+{3\over4}\,
T^{21}={7\over4}$, respectively. How many equations can be eliminated
from the system of equations using this simple equivalence test depends
on the symmetry properties (note, that even two different symmetry
properties can produce equivalent equations). This reduction scheme for
the system of linear equations can be used in all three algorithms
described above.

\section{Implementation in \Mathematica}

In order to check the performance and cross-check the results all three
algorithms A, B, and C were implemented in \Mathematica\ with the idea
that the best one should be incorporated into the package \EinS\ for
calculations with indexed objects \cite{Klioner:1998,Klioner:2003}. The
main parts of the implementation are:

\begin{enumerate}

\item[(1)] two routines {\tt DefObject} and {\tt
DefSymmetries} allowing one to define objects with arbitrary symmetries,

\item[(2)] routine {\tt ConstrainComponents} which explicitly generates and
solves the linear equations for individual components induced by the
symmetries for some fixed $k$,

\item[(3)] routine {\tt GuessPolynomial} implementing algorithm A by
calling
{\tt ConstrainComponents} for a sufficient number of different values
of $k$ to check if a polynomial of degree $n$ or less can be fitted to
the results,

\item[(4)] routine {\tt
CountIndependentComponents} implementing algorithm B by calling {\tt
ConstrainComponents} for $k=n$ and analyzing the resulted independent
components to compute $f(k)$ from (\ref{f(k)-f(1)-...-f(n)}), and

\item[(5)] routine {\tt ListIndependentComponents} implementing algorithm C
and providing for any $k$ both $f(k)$ and, if requested, a list for the
independent components and the dependence of the other components.

\end{enumerate}

\noindent
All the routines allow one to control all the steps of the
corresponding algorithms and, if desired, provide the user with various
additional information. The implementation consists of about 1000
lines of \Mathematica\ top level code and is available form the author
upon request.

To give a practical example let us consider the covariant Riemann
tensor $R_{ijkl}$ with $n=4$ and with its four symmetry properties
\begin{eqnarray}
&&
R_{ijkl}=R_{klij},
\nonumber\\
&&
R_{ijkl}=-R_{jikl},
\nonumber\\
&&
R_{ijkl}=-R_{ijlk},
\nonumber\\
&&
R_{ijkl}+R_{iljk}+R_{iklj}=0.
\end{eqnarray}

\noindent
The well-known result \cite{Landau:Lifshits:1972} for the covariant
Riemann tensor is $f(k)={1\over 12}\,k^2\,(k^2-1)$.

Algorithm A ({\tt GuessPolynomial}) explicitly computes the independent
components subsequently for $k=1$, $k=2$, \dots, $k=6$ to verify that a
polynomial of degree $4$ or less really fits the results. For example,
for $k=6$ this requires solving of a system of $4\times 6^4=5184$
linear equations (2526 distinct ones) with for 1296 unknowns. This is
quite a heavy task even for a modern PC (note that the system is
underdeterminated and should be solved exactly).

Algorithm B ({\tt CountIndependentComponents}) requires solving a total
of $4\times4^4=1024$ linear equations (504 distinct ones) for $4^4=256$
unknowns among which 112 components turn out to be zero and another 124
turn out to be functions of 20 independent components: $R_{2121}$,
$R_{3121}$, $R_{3131}$, $R_{3221}$, $R_{3231}$, $R_{3232}$, $R_{4121}$,
$R_{4131}$, $R_{4141}$, $R_{4221}$, $R_{4231}$, $R_{4232}$, $R_{4241}$,
$R_{4242}$, $R_{4231}$, $R_{4331}$, $R_{4332}$, $R_{4341}$, $R_{4342}$,
$R_{4343}$. This list allows one to conclude that $f(1)=0$, $f(2)=1$,
$f(3)=6$, and $f(4)=20$. These set of values allows one to
compute  the above-mentioned result for $f(k)$ directly from
(\ref{f(k)-f(1)-...-f(n)}).

Algorithm C ({\tt ListIndependentComponents}) required solving of
$P(4)=5$ systems of linear equations:

\begin{enumerate}
\item[(1)] a system of 4 equations (only
one distinct equation) with 1 unknown (corresponding to
$Y^1_1=(4,\dots)$),
\item[(2)] one system of 16 equations (9 distinct ones)
with 4 unknowns (corresponding to $Y^2_1=(3,1,\dots)$),
\item[(3)] a system of
24 equations (10 distinct ones) with 6 unknowns (corresponding to
$Y^2_2=(2,2,\dots)$),
\item[(4)] a system of 48 equations (24 distinct ones)
with 12 unknowns (corresponding to $Y^3_1=(2,1,1,\dots)$),
\item[(5)] a system
of 96 equations (44 distinct ones) with $24$ unknowns (corresponding
to $Y^4_1=(1,1,1,1,\dots)$).
\end{enumerate}

\noindent
Clearly, algorithm C produces the same
results for $f(k)$ as the other two algorithms, but requires much less
resources. This demonstrates the efficiency of algorithm C. It is
planned to include algorithm C into the next release of \EinS.

\section{Concluding Remarks}

Certainly, algorithm C can be further improved in certain cases if the
structure of particular symmetries are taken into account. Up to now
the algorithms does not account for any properties which the symmetries
may have. Here one can use the group-theoretic approach for manipulations
with indexed objects developed in
\cite{Rodionov:Taranov:1987,Ilyin:kryukov:1991,Ilyin:Kryukov:1994,Manssur:Portugal:Svaiter:2002}.
Certain further improvement could be achieved if an algorithm
to generate only distinct equations could be found.
However, it is doubtful that such an algorithm
would be computationally cheaper as the currently used algorithm to find and to drop
equivalent equations before solving the system of linear equations (see,
Section \ref{Section:reduction}).
It is interesting also to check if the
results of Section \ref{Section-compatibility} can be improved so that
the incompatibility of symmetries (\ref{symmetries}) could be seen in
an easier way.
Another interesting question is whether, for arbitrary symmetries, one
can express algorithmically the combinatorial ``finger exercises''
allowing one to derive $f(2)=1$, $f(3)=6$ and $f(4)= 20$ for the covariant
Riemann tensor as given, for example, in Section 92 of
\cite{Landau:Lifshits:1972}.

The considered form (\ref{symmetries}) of the symmetries does not allow
us to consider some important cases. For example, the definition of a
symmetric trace-free (STF) tensor required that a contraction of
$A_{i_1i_2\dots i_l}$ with the Kronecker symbol $\delta_{i_ai_b}$
vanishes for any $a$ and $b$. Such a symmetry cannot be written in the
form (\ref{symmetries}) and is out of the scope of this paper. On the
other hand, STF tensors plays very important role in modern physics
\cite{DSX,Hartmann:1994,Klioner:Soffel:2000} and it is important to have efficient
algorithms to store them and manipulate with them. It can be
demonstrated that the main results of this paper can be also used
for symmetries involving contractions with objects each component of
which has some numerical value. That is, one can consider symmetries of the form

\begin{equation}
\label{symmetries-contraction}
\sum_{j=1}^{p} a_{j}\
\sum_{i_1,i_2,\dots,i_m=1}^k
B_{i_1i_2\dots\,i_m}\ A^{\overline{\pi}_{j}}=b,
\end{equation}

\noindent
where $a_j$ and $b$ are numbers, $i_1$, $i_2$, \dots, $i_m$ are dummy
indices over which the contraction is performed, $B_{i_1i_2\dots\,i_m}$
is a number for any values of its indices (this can be, e.g. the
Kronecker $\delta_{ij}$ or the fully antisymmetric Levi-Civita symbol
$\varepsilon_{ijk}$, or anything else), and $\overline{\pi}_{j}$ is an
arbitrary permutation of $n$ indices containing $m$ dummy indices
$i_1$, $i_2$, \dots, $i_m$ and $n-m$ free ones. This case will be
treated in a separate publication.



\begin{thebibliography}{10}


\bibitem{DSX}
Damour, T., Soffel, M., and Xu, C.:
General relativistic celestial mechanics.
Physical Review D 43 (1991) 3273--3307;
Physical Review D 45 (1992) 1017--1044;
Physical Review D 47 (1993) 3124--3135

\bibitem{Comtet:1974}

Comtet, L.: Advanced Combinatorics. Reidel, Dordrecht (1974)

\bibitem{Hartley:1996}
Hartley, D.: Overview of computer algebra in relativity. In: Hehl F.,
Puntigam R., Ruder H. (eds.): Relativity and Scientific Computing.
Springer, Berlin (1996) 173--191

\bibitem{Hartmann:1994}
Hartmann, T., Soffel, M. Kioustelidis, T.:
On the use of STF-tensors in celestial mechanics.
Celestial Mechanics, 60 (1994) 139--159

\bibitem{Ilyin:kryukov:1991}
Ilyin, V., Kryukov, A.:
Symbolic simplification of tensor expressions using symmetries, dummy
indices and identities. In: Watt, S. (ed.) ISSAC'91, Proceedings of the
1991 International Symposium on Symbolic and Algebraic Computation
(Bonn, 1991), ACM Press, Singapore (1991) 224--228

\bibitem{Ilyin:Kryukov:1994}

Ilyin, V., Kryukov, A.:
A symbolic simplification algorithm for tensor expressions in computer
algebra. Programmirovanie (Computer Science), Nauka, Moscow,  (January
1994) 83--91, in Russian

\bibitem{Jukna:2001}

Jukna, S.: Extremal Combinatorics. Springer, Berlin (2001)

\bibitem{Klioner:2003}

Klioner, S.A.: \EinS. In:  Grabmeier, J., Kaltofen, E., Weispfenning,
W. (eds.), Computer Algebra Handbook: Foundations, Applications,
Systems. Springer, Heidelberg (2003) 469

\bibitem{Klioner:1998}

Klioner, S.A.: New system for indicial computation and its
applications in gravitational physics.  Computer Physics
Communications, {\bf 115} (1998) 231-244

\bibitem{Klioner:Soffel:2000}

Klioner, S., Soffel, M.: Relativistic Celestial Mechanics
with PPN Parameters. Physical Review D, 62 (2000) 024019

\bibitem{Landau:Lifshits:1972}

Landau, L., Lifshits, E.:
The Classical Theory of Fields
Pergamon Press, Oxford (1972)

\bibitem{Maccallum:1987}

MacCallum, M.: Symbolic computation in relativity theory.
In:
Davenport, J. (ed.), EUROCAL'87, European Conference on Computer
Algebra (Berlin, 1987), Springer, Berlin (1987) 34--43

\bibitem{Manssur:Portugal:Svaiter:2002}

Manssur, L.R.U., Portugal, R., Svaiter, B.F.:
Group-Theoretic Approach for Symbolic Tensor Manipulation.
International Journal of Modern Physics C, 13 (2002) 859--879

\bibitem{Rodionov:Taranov:1987}

Rodionov, A., Taranov, A.:
Combinatorial aspects of simplification of algebraic expressions.
In: Davenport, J. (ed.), EUROCAL'87, European Conference on Computer
Algebra (Berlin, 1987), Springer, Berlin (1987) 192--201

\end{thebibliography}
\end{document}